\documentclass[twocolumn,showpacs,preprintnumbers,prl,aps,amssymb,superscriptaddress]{revtex4}
\usepackage{graphicx}
\usepackage{dcolumn}
\usepackage{bm}
\usepackage{color}

\begin{document}

\newcommand{\ie}{{\it i.e.}}
\newcommand{\eg}{{\it e.g.}}
\newcommand{\etal}{{\it et al.}}


\title{Isotropic three-dimensional gap in the iron-arsenide superconductor LiFeAs from directional heat transport measurements}


\author{M.~A.~Tanatar}
\altaffiliation{e-mail: tanatar@ameslab.gov} 
\affiliation{The Ames Laboratory, Ames, Iowa 50011, USA}

\author{J.-Ph. Reid}
\affiliation{D\'epartement de physique
\& RQMP, Universit\'e de Sherbrooke, Sherbrooke, Canada}

\author{S. Ren\'e de Cotret}
\affiliation{D\'epartement de physique
\& RQMP, Universit\'e de Sherbrooke, Sherbrooke, Canada}

\author{N. Doiron-Leyraud }
\affiliation{D\'epartement de physique
\& RQMP, Universit\'e de Sherbrooke, Sherbrooke, Canada}

\author{F. Lalibert\'e}
\affiliation{D\'epartement de physique
\& RQMP, Universit\'e de Sherbrooke, Sherbrooke, Canada}

\author{E. Hassinger}
\affiliation{D\'epartement de physique
\& RQMP, Universit\'e de Sherbrooke, Sherbrooke, Canada}

\author{J.~Chang}
\affiliation{D\'epartement de physique
\& RQMP, Universit\'e de Sherbrooke, Sherbrooke, Canada}

\author{H. Kim}
\affiliation{The Ames Laboratory, Ames, Iowa 50011, USA}
\affiliation{Department of Physics and Astronomy, Iowa State
University, Ames, Iowa 50011, USA }

\author{K. Cho}
\affiliation{The Ames Laboratory, Ames, Iowa 50011, USA}

\author{Yoo~Jang~Song}
\affiliation{Department of Physics, Sungkyunkwan University, Suwon, Gyeonggi-Do 440-746, Republic of Korea}

\author{Yong~Seung~Kwon}
\affiliation{Department of Physics, Sungkyunkwan University, Suwon, Gyeonggi-Do 440-746, Republic of Korea}

\author{R.~Prozorov}
\affiliation{The Ames Laboratory, Ames,
Iowa 50011, USA} \affiliation{Department of Physics and Astronomy,
Iowa State University, Ames, Iowa 50011, USA }

\author{Louis Taillefer}
\altaffiliation{e-mail: louis.taillefer@physique.usherbrooke.ca }
\affiliation{D\'epartement de physique
\& RQMP, Universit\'e de Sherbrooke, Sherbrooke, Canada}
\affiliation{Canadian Institute for
Advanced Research, Toronto, Ontario, Canada}

\date{\today}


\begin{abstract}
The thermal conductivity $\kappa$ of the iron-arsenide superconductor LiFeAs ($T_c \simeq$ 18~K) 
was measured in single crystals at temperatures down to $T \simeq 50$~mK and in magnetic fields up to $H = 17$~T, 
very close to the upper critical field $H_{c2} \simeq 18$~T.
For both directions of the heat current, 
parallel and perpendicular to the tetragonal $c$ axis,
a negligible residual linear term $\kappa/T$ is found as $T \to 0$, revealing that there are no zero-energy quasiparticles in the superconducting state. 
The increase in $\kappa$ with magnetic field is the same for both current directions and it follows closely the dependence expected for an isotropic superconducting gap. There is no evidence of multi-band character, whereby the gap would be different on different Fermi-surface sheets.
These findings show that the superconducting gap in LiFeAs is isotropic in 3D, without nodes or deep minima anywhere on the Fermi surface.
Comparison with other iron-pnictide superconductors suggests that a nodeless isotropic gap is a common feature at optimal doping (maximal $T_c$).

\end{abstract}


\maketitle


Because the structure of the superconducting gap as a function of direction reflects the pairing interaction, it can shed light on the nature of the pairing mechanism.
In the iron pnictides, the experimental situation in this respect remains unclear and so far suggests the lack of any universal picture. 
Several studies agree on the existence of nodes in the superconducting gap of the low-$T_c$ materials KFe$_2$As$_2$ \cite{Shiyan,HashimotoKpure} and LaFePO \cite{lapfo1, lapfo2,lapfo3}. In BaFe$_2$As$_2$-based superconductors, signatures of nodal behavior were observed in heavily K-doped samples \cite{Fukazawa} and in P-doped compounds \cite{HashimotoP}, while in Co- and Ni-doped compounds the superconducting gap shows nodes only away from optimal doping (maximal $T_c$) \cite{Martin3D,TanatarPRL2010,Reid3D}. 

%
The material LiFeAs may prove important in the study of iron-based superconductivity 
because it is stoichiometric, and so can in principle be made with low levels of disorder,
and it has a relatively high $T_c$.
The Fermi surface of this material has four (or five) sheets, two electron pockets centered near the $M$-point of the Brillouin zone, and two (three) hole pockets centered around the $\Gamma$-point \cite{Sadovskii}.  
ARPES measurements for $k_z$=0 found an isotropic in-plane superconducting gap whose magnitude on the electron sheets, $\Delta _e$, is approximately two times larger than on the hole sheets, $\Delta _h$ \cite{BorisenkoPRL}. 
Specific heat \cite{BorisenkoSpecHeat}, penetration depth \cite{HyunsooLiFeAs,microwaveLiFeAs} and lower critical field \cite{LiFeAsHc1} measurements
were interpreted in terms of a fully isotropic, $k$-independent gap $\Delta(k)$, with $\Delta_e \simeq 2 \Delta_h$.
However, none of these studies has directional resolution  to locate out of plane nodes, such as three dimensional (3D) nodes found in the under- and over-doped Co-Ba122 \cite{Reid3D,Hirschfeld3D}


%

In this Letter, we report a study of the 3D superconducting gap structure of LiFeAs using thermal conductivity,  
a bulk probe used previously to locate directions of gap nodes in heavy-fermion \cite{CeIrIn5} and iron-pnictide \cite{Reid3D} superconductors. 
We found that for directions of heat flow parallel and perpendicular to the tetragonal $c$ axis the thermal conductivity of LiFeAs
closely follows expectations for a single isotropic superconducting gap, with no evidence of nodes or deep minima in any direction on any part of the Fermi surface. 
%
%


{\it Experimental.--}
Single crystals of LiFeAs were grown in a sealed tungsten crucible using a Bridgeman method \cite{Song2010},
and stored in sealed ampoules. 
Immediately after opening the ampoules, samples for in-plane resistivity, Seebeck and thermal conductivity measurements were 
cleaved and shaped into parallel bars $(1-2) \times (0.3-0.5) \times (0.05-0.1)$ mm$^3$ ($a \times b \times c$). 
Silver wires were soldered to the samples \cite{SUST}, yielding low-resistance contacts ($\simeq 100~\mu \Omega$). 
Samples for inter-plane resistivity and thermal conductivity, 
with dimensions $(0.5-1) \times (0.5-1) \times (0.1-0.3)$ mm$^3$,
were measured using a two-probe technique \cite{TanatarPRB2009,Reid3D}. , with contacts covering the whole $ab$-plane area of the sample. 
After contacts were made, the samples were covered with Apiezon N grease to prevent oxidation.

The thermal conductivity $\kappa$ was measured in a standard one-heater-two thermometer technique.
In both in-plane ($\kappa_a$) and inter-plane ($\kappa_c$) heat transport measurements, the magnetic field $H$ was applied along the [001] tetragonal $c$ axis. 
Measurements were done on warming after cooling in a constant field from above $T_c$, to ensure a homogeneous field distribution. 
%


\begin{figure}
\centering
\includegraphics[width=8.5cm]{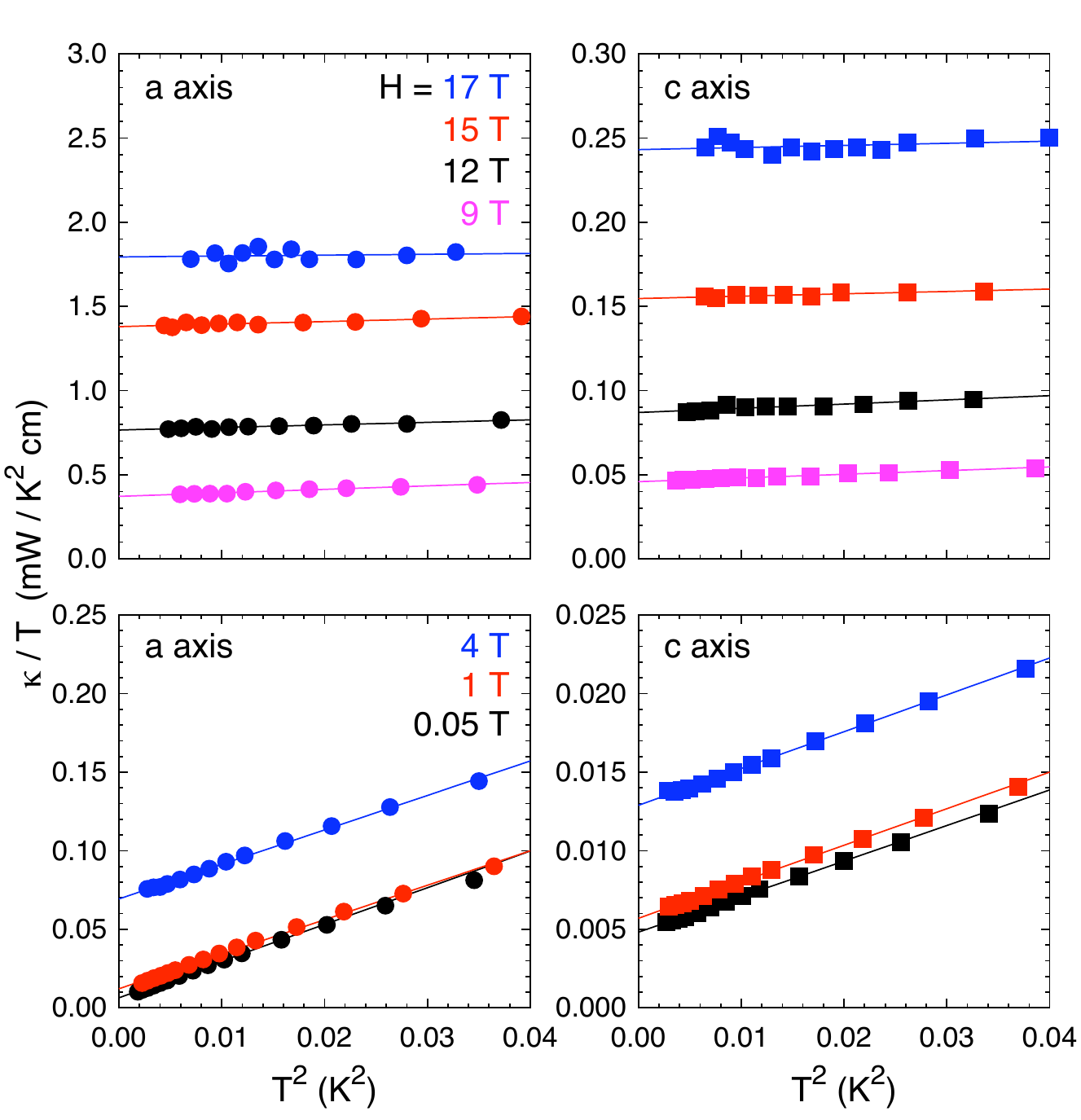}
\caption{\label{T2field} 
Thermal conductivity of LiFeAs as a function of temperature, plotted as $\kappa/T$ vs $T^2$, 
for a heat current in the basal plane (left panels) 
and along the tetragonal $c$ axis (right panels), 
measured for different values of the magnetic field $H$ as indicated.
Solid lines are linear fits used to extrapolate the residual linear term $\kappa_0/T$ at $T = 0$, 
plotted vs $H$ in Fig.~2. 
}
\end{figure}


%
{\it Temperature dependence.--}
The thermal conductivity of LiFeAs is displayed in Fig.~\ref{T2field}, for different magnetic fields up to 17~T.
The linear fits show that the data below 0.2 K are well described by the function 
$\kappa/T = a + b T^2$. The first term, $a \equiv \kappa_0 /T$, is the residual linear term, entirely due to electronic excitations \cite{NJP2009}. 
The second term is due to phonons, which at low temperature are scattered by the sample boundaries.

The magnitude of the residual linear term is extremely small. 
For both directions of heat flow, $\kappa_{0}/T \simeq 5~\mu$W / K$^2$ cm.  
These values are within the absolute accuracy of our measurements, approximately $\pm$~5~$\mu$W / K$^2$ cm \cite{Li2008,Boaknin2003}.
Therefore, our LiFeAs samples exhibit a negligible residual linear term for both in-plane and inter-plane directions. 
Comparison with the normal-state conductivity $\kappa_{\rm N}/T$, estimated using the Wiedemann-Franz law -- 
$\kappa_{\rm N}/T = L_0 / \rho_0$ where $L_0 \equiv (\pi^2/3)(k_{\rm B}/e)^2$ --
applied to the extrapolated residual resistivity $\rho_0$ (see Fig.~3), 
as discussed in Ref.~\onlinecite{Reid3D},
gives a ratio $(\kappa_0/T) / (\kappa_{\rm N}/T) \simeq$ 1 \% (0.1~$\%$) for flow parallel (perpendicular) to the $c$ axis.

These $\kappa_0/T$ values are much smaller than theoretical expectation for a nodal superconductor
(for a gap without nodes, $\kappa_0/T = 0$ \cite{NJP2009}).
For a quasi-2D $d$-wave gap, with four line nodes along the $c$ axis, the residual linear term is given, in the clean limit, 
by
$\kappa_{0}/T = (k^2_{\rm B}/6 c) (k_{F} v_{F} / \Delta_{0})$,
where $c$ is the interlayer separation, $k_F$ and $v_F$ the 
Fermi wavevector and velocity at the node, respectively, and $\Delta_0$ the gap maximum \cite{Graf1996,Durst2000,Hawthorn2007,NJP2009}.
Taking $c$ = 6.36~$\AA$, $v_F$ = 1~eV~$\AA$ = $1.5 \times 10^5$ m/s \cite{BorisenkoPRL,Kordyuk}, and a typical Fermi wavevector 
for electron sheets of the Fermi surface, $k_F$ = 0.2~$(\pi/a)$ = 0.16~\AA$^{-1}$ \cite{Borisenko10085234},
we get $\kappa_{0}/T \simeq 140~\mu$W / K$^2$ cm, assuming a weak-coupling $\Delta_0 = 2.14$ $k_{\rm B} T_c$, 
not far from the experimentally determined gap \cite{Inosov}.
This is at least 20 times larger than the value extracted from our fits to the $\kappa/T$ vs $T$ data.
In those materials where universal heat transport has been verified, proving the presence of a line node in the gap, 
the measured value of $\kappa_0/T$ is in good quantitative agreement with theoretical expectation
\cite{Taillefer1997,Hawthorn2007,Suzuki2001,CeIrIn5}. 
Thus we can safely conclude that the gap in LiFeAs does not contain a line of nodes anywhere on the Fermi surface. 
Importantly, the fact that $\kappa_0/T \simeq 0$ for both $\kappa_a$ and $\kappa_c$ rules out not only vertical
but also horizontal line nodes, including those away from the $k_z=0$ plane.


\begin{figure}
\centering
\includegraphics[width=8.5cm]{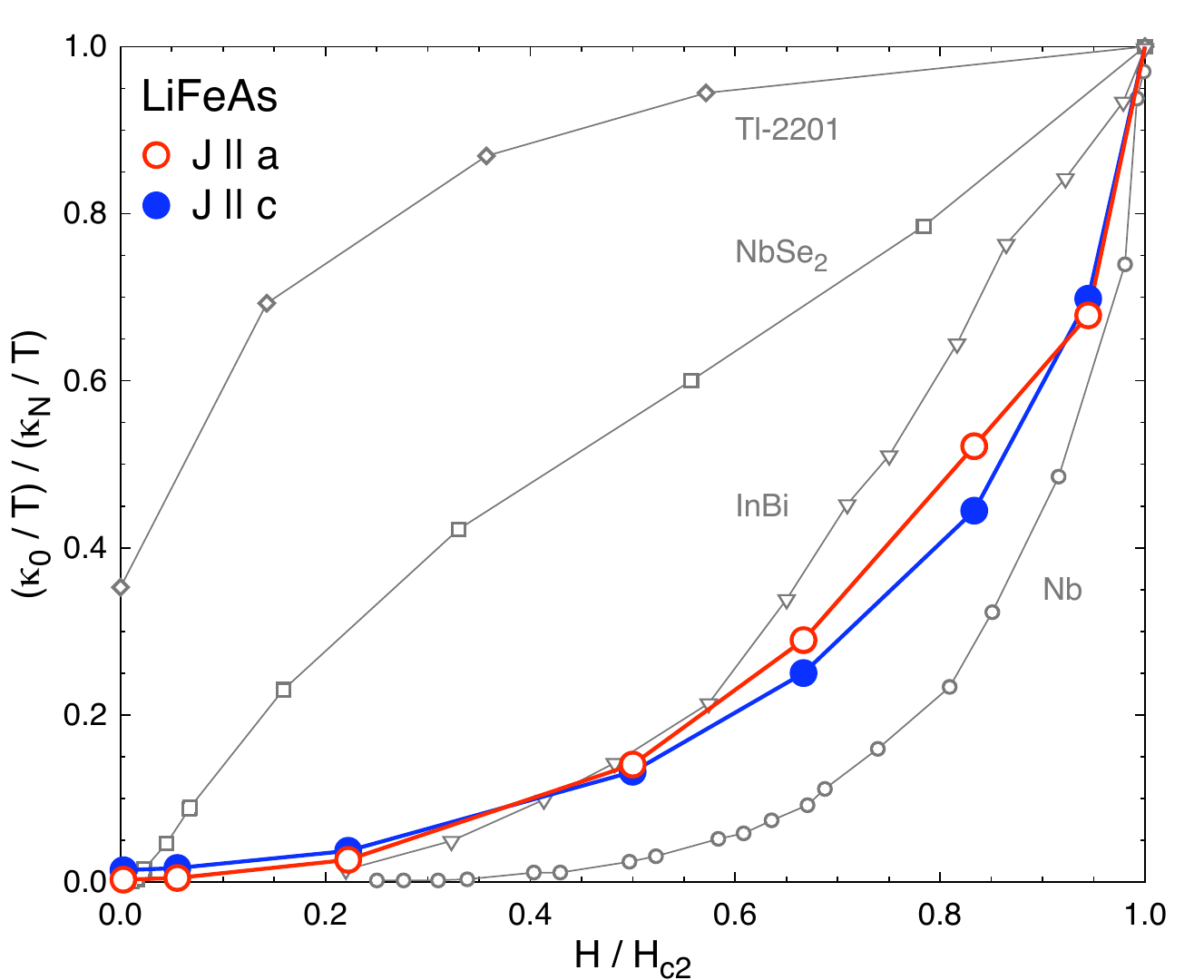}
\caption{\label{kappavsHovHc2}
Residual linear term $\kappa_0/T$ in the thermal conductivity of LiFeAs as a function of magnetic field $H$ (applied along the tetragonal $c$ axis), 
plotted on scales normalized to the normal state.
$\kappa_{\rm N}/T$ is the normal-state conductivity estimated from the Wiedemann-Franz law (see text);
$H_{c2}$ is the upper critical field in the $T=0$ limit (see Fig.~3).
The same field dependence is observed for the two directions of heat flow, along ($J~||~c$) and perpendicular ($J~||~a$) to the $c$ axis.
This isotropic behaviour is very similar to that of standard isotropic $s$-wave superconductors, 
as in the clean Nb and the dirty InBi shown here (reproduced from Ref.~\onlinecite{Li2007}).
For comparison, we also reproduce data for the $d$-wave (nodal) superconductor Tl-2201 
\cite{Proust2002} and the multi-band $s$-wave superconductor NbSe$_2$ 
\cite{Boaknin2003}.
}

\end{figure}


{\it Field dependence.--}
Our zero-field data show that there are no zero-energy quasiparticle excitations in LiFeAs, and therefore no nodes in the gap structure 
anywhere on the Fermi surface.
By applying a magnetic field, we can now investigate quasiparticles at energies above zero. 
In a type-II $s$-wave superconductor, a field applied perpendicular to the heat flow promotes heat transport
by allowing tunneling between the quasiparticle states localized in the core of adjacent vortices \cite{Golubov2011}.
The stronger the field, the closer the vortices, exponentially favouring the tunneling process, 
controlled by the ratio of coherence length $\xi_0$ to inter-vortex separation \cite{Golubov2011,Boaknin2003}.
For a full isotropic gap, this yields an exponential growth in $\kappa$ vs $H$, as shown in Fig.~2 for Nb.
Now if the gap is depressed on some region of the Fermi surface -- either by being smaller on one sheet (multi-band character) or by having a strong
angle dependence leading to a deep minimum in some $k$ direction (gap anisotropy) -- the tunneling will be enhanced, 
since $\xi_0 \propto v_{\rm F} / \Delta_0$ will be longer.
This in turn will enhance the thermal conductivity at low field, as observed for example in the multi-band $s$-wave superconductor 
NbSe$_2$ \cite{Boaknin2003} (see Fig.~2), 
or in the highly anisotropic $s$-wave superconductor LuNi$_2$B$_2$C \cite{Boaknin2001}.

In Fig.~2, we show the field dependence of $\kappa_0/T$ in LiFeAs, obtained by extrapolating the in-field $\kappa/T$ vs $T$ data of Fig.~1.
Both axes of the plot are normalized to the respective normal-state value.
$\kappa_0/T$ is measured relative to the normal-state residual conductivity $\kappa_{\rm N}/T = L_0 / \rho_0$, 
with the residual resistivity $\rho_0$ obtained by extrapolating $\rho(T)$ to $T=0$ (see Fig.~3).
Note that in the ratio $(\kappa_0/T)/(\kappa_{\rm N}/T)$ the usual uncertainties in the geometric factors of the samples cancel out, since heat and charge transport are measured using the same contacts. 
The only uncertainty lies in the $T=0$ extrapolation of $\kappa/T$ to get $\kappa_0/T$ (well below $\pm$~10~$\%$; see Fig.~1)
and of $\rho(T)$ to get $\rho_0$ (of order $\pm$~20-30~$\%$; see Fig.~3). 
The field axis in Fig.~2 is measured relative to the $T=0$ upper critical field $H_{c2}(0) \simeq 18$~T, obtained by smoothly extrapolating 
$H$ vs $T_c$ data to $T_c = 0$, where $T_c$ is detected in thermopower measurements on LiFeAs (see Fig.~3).
The value $H_{c2}(0) \simeq 18$~T is consistent with tunnel-diode-resonator measurements on the same batch of crystals \cite{LiFeAsTDR}.


\begin{figure}
\centering
\includegraphics[width=8.5cm]{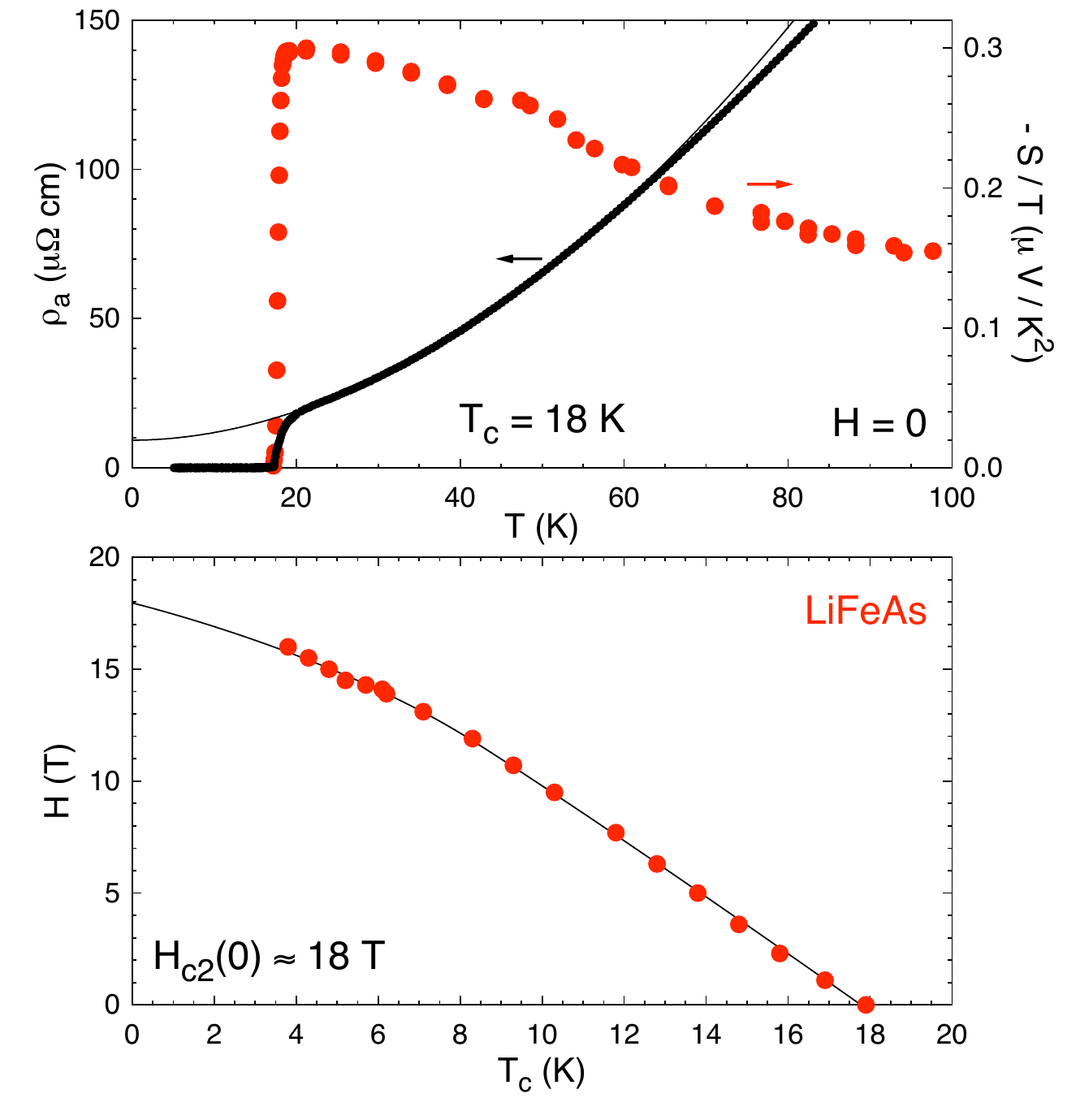}
\caption{\label{thermopower} 
Top panel: 
In-plane resistivity ($\rho_a$; small black dots) 
and Seebeck coefficient (thermopower) ($S$; large red dots) of LiFeAs, 
measured in zero magnetic field $H$, plotted as $-~S/T$ vs $T$.
Both give a zero-field superconducting transition temperature $T_c = 18$~K.
The line is a quadratric fit to the $\rho_a(T)$ data below 50~K, extended to $T=0$
in order to extract an extrapolated value of the normal-state residual resistivity 
$\rho_0 \simeq 10~\mu \Omega$~cm. 
The negative value of $S$ indicates that electron-like carriers dominate the conductivity
of LiFeAs at low temperature.
Bottom panel: 
Temperature dependence of the superconducting upper critical field $H_{c2}(T)$, determined by
detecting $T_c$ in $S/T$ vs $T$ for different field strengths.
The line is a smooth extrapolation to $T=0$, giving an estimate of the zero-temperature
critical field: $H_{c2}(0) \simeq 18$~T. 
}
\end{figure}


In Fig.~2, the field dependence of $\kappa_0/T$ in LiFeAs is seen to be isotropic, slow at low $H$ and rapid as $H$ approaches $H_{c2}$.
This upward curvature of $\kappa_0/T$ vs $H$ is typical of isotropic $s$-wave superconductors like Nb (clean limit) and InBi (dirty limit), 
as shown in Fig.~2. 
It is opposite to the field dependence expected for a gap with nodes \cite{NJP2009}, 
as illustrated in Fig.~2 with data for the $d$-wave cuprate superconductor Tl-2201 \cite{Proust2002}.
In this case, the Doppler shift of delocalized quasiparticle excitations (not confined to the vortex cores) yields a rapid initial rise \cite{Hirschfeld-vortex}.

The in-field data not only confirms the absence of nodes in the gap of LiFeAs, it also shows that the gap is isotropic in 3D, the same in and out of the basal plane.
Importantly, there is no evidence of any suppression of the gap in some direction or on some sheet of the Fermi surface. 
Indeed, as far as the quasiparticle transport is concerned, the superconducting gap appears to have the same uniform value everywhere on the Fermi surface.

{\it Relation to multi-band scenario.--}
The slow rise of $\kappa_0/T$ at low $H$ in LiFeAs is very different from the rapid rise seen in typical multi-band superconductors such as
MgB$_2$ \cite{Sologubenko} and NbSe$_2$ \cite{Boaknin2003} (see Fig.~2), 
in which the magnitude of the $s$-wave superconducting gap is significantly different on two sheets of the Fermi surface.
In both MgB$_2$ and NbSe$_2$, the small gap is roughly one third of the large gap, which translates into the existence of a field scale $H^\star \simeq H_{c2}/9$ sufficient to suppress superconductivity on the small-gap Fermi surface, 
which can then contribute its full normal-state conductivity even deep inside the vortex state \cite{Golubov2011,Boaknin2003}. 
Specifically, at $H = H_{c2}/5 > H^\star$, $\kappa_0/T$ is already half (one third) of $\kappa_{\rm N}/T$ in MgB$_2$ (NbSe$_2$). 
If the gap on the electron Fermi surface of LiFeAs were 2 to 3 times larger than the gap on the hole Fermi surface, as reported by ARPES studies \cite{BorisenkoPRL},
we would expect a significant enhancement of $\kappa_0/T$ on a field scale $H^\star \simeq H_{c2}/9 - H_{c2}/4$.
No such enhancement is observed.

Two effects could possibly reconcile the small value of $\kappa_0/T$ at low $H$ in LiFeAs with a small gap on the hole Fermi surface.
The first derives from the fact that it is not the gap $\Delta$ that controls the tunneling, 
and hence the heat transport, but the coherence length $\xi_0 \propto v_{\rm F} / \Delta$ \cite{Golubov2011}.
A small value of $v_{\rm F}$ on the hole surface could indeed compensate for the smaller gap. 
Specifically, if $v_{\rm F}^e / v_{\rm F}^h = \Delta_e / \Delta_h$, then $\xi_e = \xi_h$ and no multi-band feature in the $H$ dependence of 
$\kappa_0/T$ is expected. 
%
%
ARPES data does suggests that $v_{\rm F}^e > v_{\rm F}^h$ \cite{BorisenkoPRL,BorisenkoSpecHeat} and it may be that $\xi_e \simeq \xi_h$ in LiFeAs.
This would make $H^\star \simeq H_{c2}$.

The second effect is if the normal-state conductivity of the hole Fermi surface were much smaller than that of the electron surface, 
{\it i.e.} if $\sigma_h << \sigma_e$, or $\kappa_{\rm N}^h/T << \kappa_{\rm F}^e/T$.
The relative contribution of the small-gap hole Fermi surface at low $H$ would then be a small fraction of the 
total $(\kappa_0/T)/(\kappa_{\rm N}/T)$ and hence difficult to resolve.
Empirical evidence that $\sigma_h < \sigma_e$ in LiFeAs comes from the fact that both Hall \cite{Hall-LiFeAs}
and Seebeck (Fig.~3) coefficients are negative at low temperature.


{\it Comparison to other pnictides.--}
%
%
LiFeAs exhibits a temperature dependence of resistivity and a pressure dependence of $T_c$ that are consistent with an effective doping level close to optimal
(where $T_c$ is maximal). 
Now, at optimal doping, both Co-Ba122 and K-Ba122 show a full isotropic gap in 3D \cite{TanatarPRL2010,Reid3D,Luo2009,Reid2011}, 
just as reported here for LiFeAs.
By contrast, in the low-$T_c$ stoichiometric superconductors KFe$_2$As$_2$ \cite{Shiyan} and LaFePO \cite{lapfo1,lapfo2,lapfo3} the 
superconducting gap has nodes. 
This suggests that there may be a correlation between a high $T_c$ and a full, isotropic, nodeless gap.
In other words, high-temperature superconductivity in iron-based materials would appear to thrive on an isotropic gap, in contrast
with high-temperature superconductivity in copper oxides, which is intrinsically anisotropic and nodal. 
The only compound which shows nodal behavior at optimal doping is BaFe$_2$(As$_{1-x}$P$_{x}$)$_2$ \cite{HashimotoP}. 
It remains to be seen whether this may be due to some unique feature of the multi-sheet Fermi surface in that material, 
such as a more pronounced $c$-axis dispersion.


{\it Summary.--}
Our directional measurements of quasiparticle transport in the $T=0$ limit 
show that the superconducting gap of LiFeAs is nodeless and isotropic in all directions.
This excludes $d$-wave symmetry, and any other symmetry that requires line nodes on any piece of the multi-sheet Fermi surface of this superconductor.
Symmetries consistent with this constraint include $s$-wave and $s_{\pm}$ 
(whereby a full gap changes sign from the electron Fermi surface to the hole Fermi surface \cite{Mazin2010}). 
A nodeless isotropic gap is also found in the iron-pnictide superconductors Co-Ba122 \cite{Reid3D} and K-Ba122 \cite{Luo2009,Reid2011} at optimal doping, 
suggesting a possible connection between isotropic gap and maximal $T_c$.


Work at The Ames Laboratory was supported by the Department of Energy-Basic Energy Sciences under Contract No. DE-AC02-07CH11358.
R.P. acknowledges support from the Alfred P. Sloan Foundation.
Y.S.K. acknowledges support from Basic Science (No. 2010-0007487) and Mid-career (No. 2010-0029136) Researcher Programs through NRF grant funded by MEST.
L.T. acknowledges support from CIFAR, NSERC, CFI, FQRNT and a Canada Research Chair.


\end{document}